\documentclass[aps,prc,showpacs,nofootinbib,twocolumn]{revtex4}
\usepackage{epsfig}
\usepackage{graphicx,amsmath}
\newcommand\nn{\nonumber}

\def\phi{\varphi}

\newcommand\ba{\begin{eqnarray}}
\newcommand\ea{\end{eqnarray}}
\newcommand\be{\begin{equation}}
\newcommand\ee{\end{equation}}

\begin{document}

\title{Measuring the deviation from the Rutherford formula}
\author{E.~A.~Kuraev}
\email{kuraev@theor.jinr.ru}
\affiliation{\it JINR-BLTP, RU-141980 Dubna, Moscow region, Russian Federation}
\author{M.~Shatnev}
\affiliation{\it National Science Centre "Kharkov Institute of Physics and Technology", 61108 Akademicheskaya 1, Kharkov, Ukraine}
\author{E.~Tomasi-Gustafsson}
\email{etomasi@cea.fr}
\affiliation{\it CEA,IRFU,SPhN, F-91191 Gif-sur-Yvette, and  \\
Institut de Physique Nucl\'eaire, CNRS/IN2P3 and Universit\'e  Paris-Sud, France}

\date{\today}

\begin{abstract}
Modern experiments with heavy ion-leptons collisions open the possibility to
measure the deviation of cross section of small angles electron(positron)-ion
elastic scattering from the Rutherford formula due to multiple virtual photons exchange. The charge asymmetry and the polarization of the scattered leptons are calculated and numerical predictions are given. A generalization to elastic proton-nucleus scattering is discussed.
\end{abstract}

\maketitle

%\section{Introduction}

Elastic, inelastic and deep-inelastic lepton scattering on hadrons is considered the most precise way to get information about the internal structure of the hadron. An elegant formalism has been derived for those reactions, assuming that the interaction proceeds through the exchange of one photon. Lowest order QED calculations are justified by the smallness of the electromagnetic fine structure constant, $\alpha=e^2/4\pi=1/137$. However, for a heavy target of charge $Z$, the expansion parameter is not $\alpha$, but $Z\alpha$. It may be sizable and higher orders should be taken into account. The motivation of this Brief Report is to discuss high order corrections, due to multiple photon exchange, to the cross section and to polarization observables in elastic lepton-(heavy) ion scattering. The expression of the correction to the polarization of the scattered electron is derived. These predictions can be experimentally tested. If deviations from the Born expectations will be measured in dedicated experiments, care must be taken in the interpretation of the experimental data, in particular concerning the properties of the hadron structure.

Elastic lepton scattering in the Coulomb field of a nucleus has been calculated by Mott \cite{Mott} who showed that the amplitude differs from the Born approximation by a phase, which depends on the lepton charge and cancels in the cross section. If one takes into account the hadron structure, then Coulomb effects appear in the cross section, too. Recently this problem has regained interest in the literature, due to the increased precision which can be achieved in the experiments \cite{Ko01}.

The lowest order correction to the cross section of relativistic electron(positron) scattering on a heavy point-like target of charge $Z$, was first derived in Refs. \cite{p1} in the middle of the previous century. The result for the cross section has the form:
\ba
\frac{d\sigma^\pm}{dO}&=&\frac{d\sigma_R}{dO}\left [ 1\pm \pi\alpha Z
\sin(\theta/2)\right  ],~\nn\\
\frac{d\sigma_R}{dO}&=&{(Z\alpha)^2}{4E^2\sin^4(\theta/2)}
\label{eq:eq1}
\ea
where $E$ is the incident energy and $\theta$ is the angle of the scattered electron in the laboratory (Lab) system. 

Two(or more) photon exchange induce a nonzero imaginary part in the scattering amplitudes, therefore the scattered electron may be polarized also in the case of unpolarized particles collisions. Let us define $\vec{e}$ the electron polarization vector. In the case of high energies and small scattering angles it can be written as:
\be
\vec{e}=\frac{2Z\alpha m}{E}\sin^3\frac{\theta}{2}
\ln\sin\frac{\theta}{2}\vec{\nu},
\label{eq:eqpol}
\ee
where
$\vec{n},\vec{n}^{'},\vec{\nu}=\vec{n}\times\vec{n}^{'}/\sin\theta$ are the unit vectors along the momenta of initial electron, scattered
electron and the normal to the scattering plane, respectively 
($\cos\theta=\vec n\cdot\vec n'$), $m$ is the electron mass. 

The contributions of higher orders of perturbation theory, i.e.
the terms of order $(Z\alpha)^n$, $n>2$ were derived during the years 1975-1979, in a series of papers \cite{p2,p3,p4}.

%\section{Appendix}
In the eikonal approximation, the elastic scattering amplitude of high energy electrons (positrons) in the Coulomb field  has the form \cite{AB81}:
\ba
f(q)=-iE\int\limits_0^\infty \rho d\rho J_0(q\rho)e^{i\kappa(\rho)},
\label{eq:eqa1}
\ea
where $\rho$ is the impact parameter, $J_0(z)$ is the Bessel function, and
$\kappa(\rho)=\kappa_0(\rho)+\kappa_1(\rho)$ is the eikonal phase. In the first
and second approximation we have
\ba
\kappa_0(\rho)=-\int_{-a}^a dt V(\rho,t); \nn
\kappa_1(\rho)=-\displaystyle\frac{\rho^2}{E}\int_{-\infty}^\infty dt\displaystyle\frac{\partial}{\partial \rho^2} V^2(\rho,t),
\label{eq:eqa2}
\ea
where $V(\rho,t)=x e/|e|\sqrt{\rho^2+t^2}$, $x=Z\alpha$, and $a$ is the
regularizing parameter of the Coulomb potential $a>>\rho$.
Moreover, it is implied that the 
energy of the electron, $E$, is large compared to its mass. A straightforward calculation leads to:
\ba
\kappa_0(\rho)=2x\ln\frac{\rho}{2a};~ \kappa_1(\rho)=\frac{\pi x^2}{2\rho E}.
\label{eq:eqa3}
\ea
Inserting (\ref{eq:eqa3}) in Eq. \ref{eq:eqa1}), we obtain for the amplitude
\ba
f(q)=-iE\int\limits_0^\infty \rho d\rho J_o(q\rho)\left(\displaystyle\frac{\rho}{2a}\right )^{2ix}\left (1+i\displaystyle\frac{\pi x^2}{2\rho E}\right),
\label{eq:eqa4}
\ea
where we took into account the smallness of the second order eikonal phase 
$|\kappa_1| <<|\kappa_0|$.
Further integration is performed using the relation \cite{p5}:
\ba
\int\limits_0^\infty d x J_0(q x) x^\mu=2^\mu (q)^{-\mu-1}\displaystyle\frac{\Gamma\left(\displaystyle\frac{1+\mu}{2}\right)}{\Gamma\left(\displaystyle\frac{1-\mu}{2}\right)}.
\ea
Using the relation $q=2E\sin(\theta/2)$, the result for the amplitude is:
\ba
f(q)=-\displaystyle\frac{x E}{2 q^2}(a q)^{-2ix}\displaystyle\displaystyle\frac{\Gamma(1+ix)}{\Gamma(1-ix)}\left[1-\displaystyle\frac{\pi x \sin(\theta/2)}{2}\Phi(x)\right],
\ea
with $\Phi(x)$ : 
\ba
\Phi(x)&=&\cos\phi(x)+i\sin\phi(x)=\frac{\Gamma(\frac{1}{2}+ix)\Gamma(1-ix)}
{\Gamma(\frac{1}{2}-ix)\Gamma(1+ix)}, 
\label{eq:eqcs}\\
x&=&\frac{Z\alpha}{\beta}.
\label{eq:eqz}
\ea
where $\beta$ is the velocity $v$ of the initial particle, in the laboratory system, in units of $c$: 
$\beta={v}/{c}= \sqrt{1-4{m^2}/{E}}$

\begin{figure}
\includegraphics[width=0.5\textwidth]{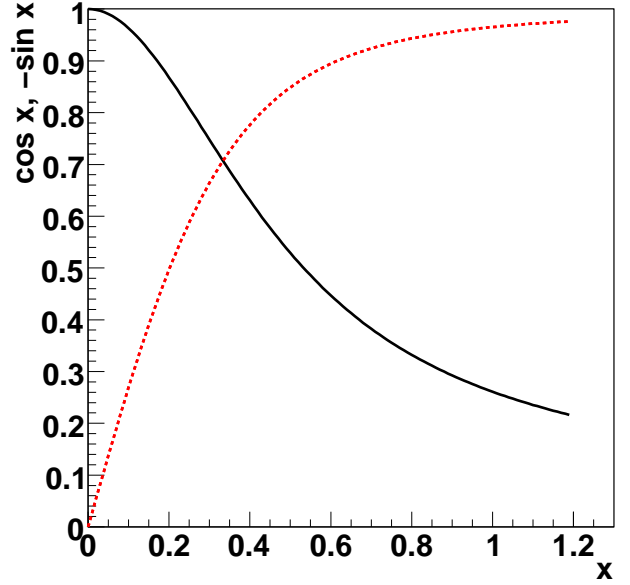}
\caption{(Color online) Dependence of the real (black, solid line) and imaginary (red, dashed line) part of $\Phi(x)$ as a function of $x$ [Eq. (\ref{eq:FDef})]. }
\label{FigReD}
\end{figure}

Using the properties of Euler gamma function \cite{p5} one obtains:
\ba
\phi(x)&=&-4\sum_{n=0}^\infty (-1)^n
\frac{x^{2n+1}}{2n+1} c_n, \nn \\
c_0&=&\ln 2; c_1=3\xi_3; ~c_2=15\xi_5,...,\nn \\ 
c_n&=&(2^{2n}-1)\xi_{2n+1},~n \geq 1.
\label{eq:phi}
\ea
Applying the Stirling formula one can write
\be
\cos(\phi(x))|_{x>>1}\sim\frac{1}{4x}.
\label{eq:st}
\ee
The functions $\cos(\phi(x))$ and $\sin(\phi(x))$ are shown in Fig.~\ref{FigReD}.

The expression for the differential cross section is:
\ba
\frac{d\sigma}{d\Omega}=\displaystyle\frac{d\sigma_R}{d\Omega}\left [1-\pi x\sin(\theta/2) \cos\phi(x)\right ].
\ea
Therefore, the effect of multiphoton exchange results in a correction to the differential cross section, which can be expressed as a multiplicative factor $\cos\phi(x)$ in front of the second term in square brackets of Eq. \ref{eq:eq1}. 

\begin{figure}
\includegraphics[width=0.5\textwidth]{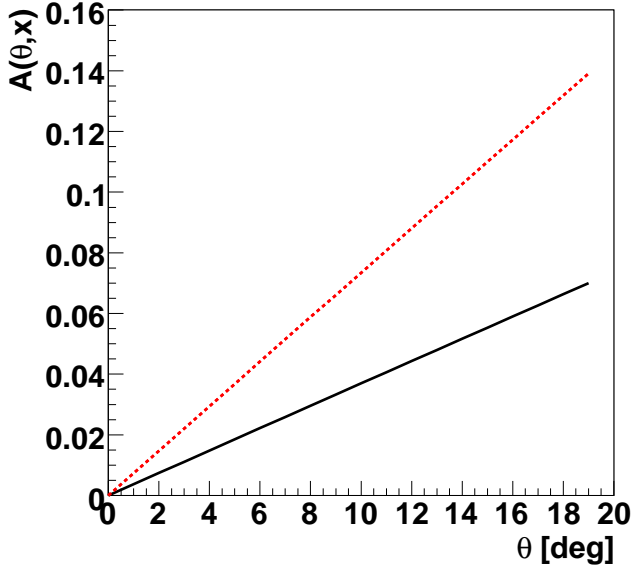}
\caption{(color online) Dependence of the charge asymmetry $A(\theta,x)$ on $\theta$, (see Eq. (\ref{eq:asym})), at $E=3$ GeV, for $x=0.15$ ($Z=20$) (black, solid line) and $x=0.6$ ($Z=82$) (red, dashed line).
}
\label{Fig:Asym}
\end{figure}

In Ref. \cite{p3} the charge asymmetry, defined as the difference of the cross sections for the scattering of electron ($\mu_-$)
and positron ($\mu_+$) on the same target of charge $Z$, was derived using the eikonal approximation:
\ba
A=\frac{\displaystyle\frac{d\sigma^{e^-Z}}{dO_-}-\displaystyle\frac{d\sigma^{e^+Z}}
{dO_+}}{\displaystyle\frac{d\sigma^{e^-Z}}{dO_-}+\displaystyle\frac{d\sigma^{e^+Z}}{dO_+}}=\pi x \sin\left (\frac{\theta}{2}\right )\cos\phi(x).
\label{eq:asym}
\ea
The charge asymmetry is shown in Fig. \ref{Fig:Asym} as a function of the polar angle $\theta$, for two different values of the ion charge: $Z=20$ (black, solid line) and $Z=82$ (red, dashed line), which correspond to $x=0.15$ and $x=0.6$, respectively.
The calculation is performed for for $E=3$ GeV, but the charge asymmetry is almost constant with energy, as long as $\beta\sim 1$. One can see that it increases with angle, and may reach measurable values of a few percent. Note that the present calculation holds for small scattering angles.

%\begin{figure}
%\includegraphics[width=0.8\textwidth]{ReD.eps}
%\caption{The dependence of function $Re D(x) = \cos(\phi(x))$ of $x$
%(see (\ref{DDefinition})).
%\label{FigReD}}
%\end{figure}

Higher order corrections induce also an effect to the polarization vector of the scattered electron. Such effect is taken into account by doing the following replacement in Eq. (\ref{eq:eqpol}):
\ba
Z\alpha\ln\sin\frac{\theta}{2} \to
F(x,\theta) = x\ln\theta+\frac{1}{4}\sin\phi(x).
\label{eq:FDef}
\ea 
 
\begin{figure}
\includegraphics[width=0.5\textwidth]{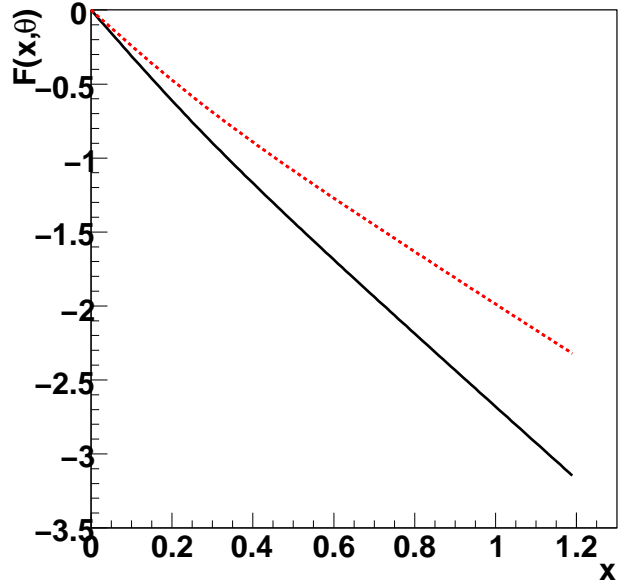}
\caption{(color online)Dependence of function $F(\theta,x)$ on $x$, Eq. (\ref{eq:FDef}), for two different values of $\theta$: $\theta= 5^{\circ}$ (black, solid line), $\theta=10^{\circ}$ (red, dashed line).}
\label{FigF}
\end{figure}

\begin{figure}
\includegraphics[width=0.5\textwidth]{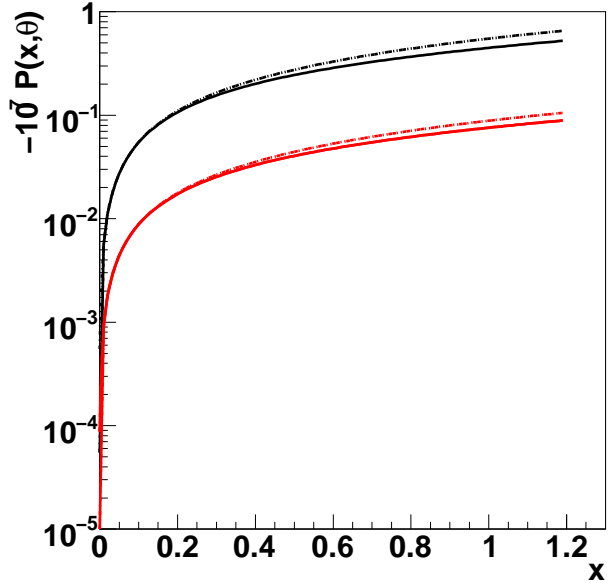}
\caption{(Color online) Dependence of the outgoing electron polarization  on $x$, (see Eqs. (\protect\ref{eq:FDef},\protect\ref{eq:eqpol})), for two different values of $\theta$: $\theta= 5$ deg (black, solid line), $\theta=10$ deg (red, solid line).}
\label{Fig:Pol}
\end{figure}

The function $F(x,\theta)$ is drawn in Fig.~\ref{FigF}.
The normal component of the polarization of the outgoing electron is drawn in Fig.~\ref{Fig:Pol} as a function of $x$ (see Eqs. (\protect\ref{eq:FDef},\protect\ref{eq:eqpol})),  for two different values of $\theta$:  $\theta= 5^{\circ}$ (black, solid line) and $\theta=10^{\circ}$ (red, dashed line). The corresponding dashed lines are the result of the first order calculation. Here we consider $E=3$ GeV, and $\beta=1$.
This observable is very small and negative (the vertical scale is multiplied by $10^7$). 
%%%%%%%%%%%%%%%%%%%%%%%%
%\section{Discussion}
%%%%%%%%%%%%%%%%%%%%%%%%

A tentative extrapolation of this formalism to hadron scattering can be suggested. Elastic peripheral scattering of high energy protons on nuclei
can be experimentally investigated at LHC. Relevant theoretical considerations have not been done, at our knowledge, and  must be performed in the frame of QCD. It is tempting to obtain a 
realistic estimation of these effects by the replacing 
\ba
x=Z\alpha \to x_c=N\alpha_s.
\ea
In the kinematical conditions when $x_c>>1$, the small angles corrected cross section is expected to have an universal $\theta$ dependence as
\ba
\frac{d\sigma}{d\sigma_R}=1+\frac{\pi}{4}\sin\left(\frac{\theta}{2}\right), \theta<<1.
\label{eq:eq9}
\ea
Similarly, the asymmetry of proton and antiproton scattering on the same target nucleus, defined as for the lepton case, is expected to be
\ba
A(p,\bar{p})=A=\displaystyle\frac {\displaystyle\frac{d\sigma^{pY(N)}}{dO_p}-\displaystyle\frac{d\sigma^{\bar{p}Y(N)}}
{dO_{\bar{p}}}}{\displaystyle\frac{d\sigma^{pY(N)}}{dO_p}+\displaystyle\frac{d\sigma^{\bar{p}Y(N)}}
{dO_{\bar{p}}}}=
\displaystyle\frac{\pi}{4}\sin\left(\displaystyle\frac{\theta}{2}\right ), \nn
\theta<<1.
\ea

In conclusion, the nontrivial behavior of the deviation from the Rutherford formula due to high order $Z\alpha$ contributions, has not yet been experimentally observed. The charge asymmetry contains the information on such deviation, it is sizable and, in principle,  measurable. The present Brief Report adds a new information, the contribution to the polarization of the scattered electron. The measurement of the degree of transverse polarization of the scattered electron (muon) gives in principle another possibility to check the importance of high order effects. However the present calculation predicts very small values.  

Such correction is particularly important in problems related to beam
monitoring and calibration for small angles lepton-nuclei scattering. It should also be considered in other processes, as very small angle Bhabha scattering, or lepton-nucleus scattering. This is also relevant to present experiments as COMPASS and HERMES. An application to hadron scattering at high energies has also been suggested.

%As was pointed out in \cite{BPS} at values of $x=Z\alpha/v\sim 1,2$ asymmetry
%practically do not depend on $Z\alpha$ and equal $A=(\pi/4)v^2\sin(\theta/2)$.

%As was pointed out in \cite{BPS} at values of $x=Z\alpha/v\sim 1,2$ asymmetry
%practically do not depend on $Z\alpha$ and equal $A=(\pi/4)v^2\sin(\theta/2)$.

%\section{acknowledgements}
One of us (E.A.K.) is grateful to V. M. Katkov for discussions and to S. Bakmaev
and Yu.M. Bystritskiy for help as well as for grant INTAS 05-1000008-8328 for financial support.

\end{document}